\documentclass[%
reprint,
superscriptaddress,
showpacs,
nobibnotes,
 amsmath,amssymb,
 aps,
prl,
floatfix,
]{revtex4-1}

\usepackage{graphicx}% Include figure files
\usepackage{dcolumn}% Align table columns on decimal point
\newcolumntype{.}{D{.}{.}{-1}}
\usepackage{bm}% bold math
\usepackage{float,array,booktabs,amsmath,amssymb}
\begin{document}

\preprint{APS/123-QED}

\title{\boldmath Benchmarking $^{136}$Xe Neutrinoless $\beta\beta$ Decay Matrix Element Calculations with the $^{138}{\rm Ba}(p,t)$ Reaction}% Force line breaks with \\
\author{B.\,M.~Rebeiro}
\email{b.rebeiro@gmail.com}
\affiliation{Department of Physics and Astronomy, University of the Western Cape, P/B X17, Bellville 7535, South Africa.}%
\author{S.~Triambak}
\email{striambak@uwc.ac.za}
\affiliation{Department of Physics and Astronomy, University of the Western Cape, P/B X17, Bellville 7535, South Africa.}%
\author{P.\,E.~Garrett}
\affiliation{Department of Physics, University of Guelph, Guelph, Ontario N1G 2W1, Canada.}%
\affiliation{Department of Physics and Astronomy, University of the Western Cape, P/B X17, Bellville 7535, South Africa.}%
\author{B.\,A.~Brown}
\affiliation{Department of Physics and Astronomy and National Superconducting Cyclotron Laboratory,
Michigan State University, East Lansing, Michigan 48824-1321, USA}
\author{G.\,C.~Ball}
\affiliation{TRIUMF, 4004 Wesbrook Mall, Vancouver, British Columbia V6T 2A3, Canada.}
\author{R.~Lindsay}
\affiliation{Department of Physics and Astronomy, University of the Western Cape, P/B X17, Bellville 7535, South Africa.}%%
\author{P.~Adsley}
\affiliation{School of Physics, University of the Witwatersrand, Johannesburg 2050, South Africa}%
\affiliation{iThemba LABS, P.O. Box 722, Somerset West 7129, South Africa}%
\author{V.~Bildstein}
\affiliation{Department of Physics, University of Guelph, Guelph, Ontario N1G 2W1, Canada.}%
\author{C.~Burbadge}
\affiliation{Department of Physics, University of Guelph, Guelph, Ontario N1G 2W1, Canada.}%
\author{A.~Diaz Varela}
\affiliation{Department of Physics, University of Guelph, Guelph, Ontario N1G 2W1, Canada.}%
\author{T.~Faestermann }
\affiliation{Physik Department, Technische Universit\"{a}t M\"{u}nchen, D-85748 Garching, Germany.}%
\author{D.\,L.~Fang}
\affiliation{Institute of Modern Physics, Chinese Academy of Sciences, Lanzhou, 730000, China}%
\affiliation{School of Nuclear Science and Technology, University of Chinese Academy of Sciences, Beijing 100049, China}
\author{R.~Hertenberger}
\affiliation{Fakult\"{a}t f\"{u}r Physik, Ludwig-Maximilians-Universit\"{a}t M\"{u}nchen, D-85748 Garching, Germany.}%
\author{M.~Horoi}
\affiliation{Department of Physics, Central Michigan University, Mount Pleasant, Michigan 48859, USA}
\author{B.~Jigmeddorj}
\affiliation{Department of Physics, University of Guelph, Guelph, Ontario N1G 2W1, Canada.}%
\author{M.~Kamil}
\affiliation{Department of Physics and Astronomy, University of the Western Cape, P/B X17, Bellville 7535, South Africa.}%
\author{K.\,G.~Leach}
\affiliation{Department of Physics, Colorado School of Mines, Golden, Colorado 80401, USA}
 \author{P.\,Z.~Mabika}
\affiliation{Department of Physics and Astronomy, University of the Western Cape, P/B X17, Bellville 7535, South Africa.}%
\affiliation{Department of Physics and Engineering, University of Zululand, Private Bag X1001, KwaDlangezwa 3886, South Africa.}
\author{J.\,C.~Nzobadila~Ondze}
\affiliation{Department of Physics and Astronomy, University of the Western Cape, P/B X17, Bellville 7535, South Africa.}%
\author{J.\,N.~Orce}
\affiliation{Department of Physics and Astronomy, University of the Western Cape, P/B X17, Bellville 7535, South Africa.}%
\author{H.\,-F.~Wirth}
\affiliation{Fakult\"{a}t f\"{u}r Physik, Ludwig-Maximilians-Universit\"{a}t M\"{u}nchen, D-85748 Garching, Germany.}%
%  \email{Second.Author@institution.edu}
% \affiliation{%
%  Authors' institution and/or address\\
%  This line break forced with \textbackslash\textbackslash
% }%
% 
% \collaboration{MUSO Collaboration}%\noaffiliation
%  \homepage{http://www.Second.institution.edu/~Charlie.Author}
% \affiliation{
%  Second institution and/or address\\
%  This line break forced% with \\
% }%
% \affiliation{
%  Third institution, the second for Charlie Author
% \affiliation{%
%  Authors' institution and/or address\\
%  This line break forced with \textbackslash\textbackslash
% }%
% 
% % 
% % \author{V.~Pesudo}%
% % %  \affiliation{University of the Western Cape, P/B X17, Bellville 7535, South Africa}
% %   \affiliation{iThemba LABS, P.O. Box 722, Somerset West 7129, South Africa}
% %  \author{B.~Singh}
% %  \affiliation{University of the Western Cape, P/B X17, Bellville 7535, South Africa}
 
\date{\today}% It is always \today, today,
 \begin{abstract}
We used a high-resolution magnetic spectrograph to study neutron pair-correlated $0^+$ states in $^{136}$Ba, produced via the $^{138}{\rm Ba}(p,t)$ reaction. In conjunction with state-of-the-art shell model calculations, these data benchmark part of the dominant Gamow-Teller component of the nuclear matrix element (NME) for $^{136}$Xe neutrinoless double beta ($0\nu\beta\beta$) decay.
% A large fragmentation of the $L = 0$ $(p,t)$ strength to excited $0^+$ states in $^{136}$Ba is observed for the first time, indicating a breakdown of the Bardeen-Cooper-Schrieffer (BCS) pairing approximation for neutrons in $^{136}$Ba.
We demonstrate for the first time an evaluation of part of a $0\nu\beta\beta$ decay NME by use of an experimental observable, presenting a new avenue of approach for more accurate calculations of $0\nu\beta\beta$ decay matrix elements. 
%Our work also demonstrates for the first time a direct estimation of part of a neutrinoless $\beta\beta$ decay nuclear matrix element (NME) using a two-nucleon transfer experiment. 
%A comparison of our experimental data with state-of-the-art shell model calculations shows that the limited model-space previously used to evaluate the NME for $^{136}$Xe neutrinoless $\beta\beta$ decay does not lead to an accurate result. On incorporating corrections due to configuration mixing with orbitals beyond the valence space, we calculate the dominant $J = 0$ Gamow-Teller part of the NME to be enhanced about a factor of 1.54 or 1.83, depending on whether the decay occurs due to light or heavy Majorana neutrino exchange. 
%The presented analysis 
%thereby presenting . %We recommend a similar analysis for other candidates.
%play an important role and are essential for the matrix element calculation in this particular case.  
%Our preliminary estimates show .    
%$^{136}$Xe neutrinoless $\beta\beta$ decay matrix element calculations.   
\end{abstract}

%\pacs{ }% PACS, the Physics and Astronomy
\maketitle
%Ever since the phenomenon of neutrino oillations~\cite{mns,pontecorvo} 
%was verified by independent experiments~\cite{Homestake,kamiokande,sno}, there has been a resurgence of activity in the field of neutrino physics. 
%The fact that neutrinos have mass results in a breakdown of the previously assumed $\gamma_5$ invariance for the weak interaction~\cite{sakurai}. 
%The massive nature of neutrinos
%~\cite{Homestake,kamiokande,sno} 
%breaks down the previously assumed $\gamma_5$ invariance for weak interactions~\cite{sakurai}. 
%Since 
%
%$\gamma_5$ invariance 
%
%weak interactions 
%massive nature of neutrinos breaks down the ~\cite{sakurai}.
The massive nature of neutrinos leads to a violation of the $\gamma_5$ invariance~\cite{sakurai} for weak interactions.
Consequently, there is substantial interest worldwide~\cite{Avignone_rmp,Bilenky:2015,Dolinski:2019} to search for standard-model-forbidden neutrinoless double beta ($0\nu\beta\beta$) decays, that violate lepton number conservation by 2 units. The observation of such decays would prove that the electron neutrino ($\nu_e$) is a Majorana fermion, and therefore indistinguishable from its antiparticle ($\bar{\nu}_e$). This is consistent with most theories beyond the standard model~\cite{Rodejohann:2011}, that attribute the smallness of neutrino masses to a violation of total lepton number at an energy scale of $\sim10^{15}$~GeV~\cite{Bilenky:2015,Rodejohann:2011}.
% 
% 
% %This has led to a resurgence of interest worldwide~\cite{Avignone_rmp,Bilenky:2015,Dolinski:2019} to search for neutrinoless double beta ($0\nu\beta\beta$) decay
% 
% %- a standard-model-forbidden process that would violate lepton number conservation by 2 units. The observation of %$0\nu\beta\beta$
%  %In such a scenario, neutrinos with definite masses ought to be Majorana particles~\cite{Rodejohann:2011,Bilenky:2015,Schechter:82}.   
% 
% %which allows the possibility for
% %neutrinoless double beta ($0\nu\beta\beta$) decays to occur. 
% 
% %The observation of such a lepton number violating process would unequivocally prove that the electron neutrino ($\nu_e$) is a Majorana fermion, and therefore indistinguishable from its antiparticle ($\bar{\nu}_e$). 
% %In fact, 
%  
% 
% %Considering the above, there are several large-scale experiments currently underway~\cite{Avignone_rmp,Bilenky:2015,Dolinski:2019}, that aim to observe $0\nu\beta\beta$ decays. 

If the mechanism driving a $0\nu\beta\beta$ decay is via the exchange of a light left-handed Majorana neutrino, then the decay amplitude is proportional to 
\begin{equation}
m_{e e} M^{0\nu}= \left|\sum_{j} |U_{ej}|^2 e^{i\alpha_j}m_j\right|M^{0\nu},
\label{eq:eff_mass}
\end{equation}
where $m_{e e}$ is the effective Majorana mass of the electron neutrino and $M^{0\nu}$ is the nuclear matrix element (NME) for the decay. The NME is expressed as the sum of Gamow-Teller~(GT), Fermi~(F) and Tensor~(T) components
\begin{equation}
M^{0\nu} = M^{0\nu}_{GT} -\left(\frac{g_V}{g_A}\right)^2 M^{0\nu}_F + M^{0\nu}_T,
\end{equation}
where the Gamow-Teller contribution is the dominant term. 
In Eq.~\eqref{eq:eff_mass}, the $U_{ej}$ are elements of the Pontecorvo-Maki-Nakagawa-Sakata (PMNS) neutrino mixing matrix~\cite{mns,pontecorvo}, the $m_j$'s are the light neutrino masses and the $\alpha_j$'s are phases in the mixing matrix. For the special case of three-neutrino mixing, the PMNS matrix is parameterized in terms of three mixing angles and one Dirac and two Majorana \mbox{CP-violating} phases~\cite{Avignone_rmp}. It is evident from Eq.~\eqref{eq:eff_mass} that in addition to the observation of a $0\nu\beta\beta$ decay process, it is also equally important to determine its half-life,
%
%decays are indeed , 
%(which would validate the Majorana nature of neutrinos), 
which would establish the absolute neutrino mass scale. Furthermore, such a measurement also has the potential to identify the correct neutrino mass spectrum~\cite{Bilenky_prd} and find extra sources of CP-violation in the leptonic sector~\cite{Pascoli:2002}. However, achieving the above (or placing stringent constraints on any new physics) requires an accurate evaluation of the NME for the decay. %This is a formidable enterprise which has
This has been at the forefront of nuclear physics research in recent times, with several approaches being used to calculate the NMEs for $0\nu\beta \beta$ decay candidates~\cite{Vergados:2016,Engel_2017}. Depending on the method used, the calculations for specific isotopes disagree with one another, differing by factors of three or more in many cases~\cite{Vergados:2016,Engel_2017}. These discrepancies result in large uncertainties for the NMEs, which not only limit the physics that can be addressed, but also the planning and execution of future $0\nu\beta\beta$ decay experiments~\cite{Engel_2017}. In contrast, the NMEs for the rare (yet standard-model-allowed) two-neutrino double beta ($2\nu\beta\beta$) decays can be extracted directly from measured  half-lives. These and other experimentally derived spectroscopic information have played a critical role in constraining various NME calculations~\cite{Horoi:2013,Freeman:2012,Frekers:2018,Pirinen:2015,Rodriguez:2010,Menendez:2009}.

One of the most promising candidates for observing $0\nu\beta\beta$ decays is~$^{136}{\rm Xe}$ $\overset{\beta\beta}{\to}$ $^{136}{\rm Ba}$. Its $2\nu\beta\beta$ decay half-life is much longer than most other cases~\cite{Saakyan}.
%Its $2\nu\beta\beta$ decay NME is determined to be $M^{2\nu} \approx 0.02~{\rm MeV}^{-1}$~\cite{Ackerman}, which is the smallest of all known candidates. 
As a result, the ratio of the $0\nu\beta\beta$ decay signal to the irreducible $2\nu\beta\beta$ decay background in the vicinity of the decay endpoint energy is expected to be larger for this particular case. In fact, a highly sensitive search for $0\nu\beta\beta$ decays was recently reported for $^{136}$Xe by the KamLAND-Zen collaboration~\cite{Gando:2016}, who placed the most stringent upper limits to date on the effective neutrino mass $m_{ee} < 61-165$~meV, depending on the choice of NME used.
\begin{table}[t]
\begin{flushleft}
\caption{Some recent evaluations of the NME for $^{136}$Xe $0\nu\beta\beta$ decay using different theoretical approaches. The calculations assume light Majorana neutrino exchange.
}
\label{tab:NMEs}
\begin{ruledtabular}
\begin{tabular}{l.}
Method &\multicolumn{1}{c}{$M^{0\nu}$}\\
 \colrule
Deformed WS-QRPA (Jilin-T\"ubingen-Bratislava)~\cite{qrpa:2018}&1.11\\
pnQRPA (Jyv\"askyl\"a)~\cite{jyvaskyla:2015}&2.91\\
Deformed Skyrme-HFB-QRPA (Chapel Hill)~\cite{qrpa_CH}&1.55\\
Spherical QRPA (Bratislava-T\"ubingen-CalTech)~\cite{Simkovic}&2.46\\
ISM (Strasbourg-Madrid)~\cite{Menendez:09}&2.19\\
ISM (Michigan)~\cite{Horoi:2013}&1.46\\
%ISM (Tokyo)~\cite{ism_tokyo}&2.28\\
CDFT~\cite{redf:2017}&4.24\\
NREDF~\cite{nredf:2013}&4.77\\
IBM-2~\cite{IBM:2015}&3.05\\
GCM~\cite{gcm:2018}&2.35\\
%\colrule
%\multicolumn{2}{c}{Weighted mean} &-3.65(6)&$-3.76(44)^c$
\end{tabular}
\end{ruledtabular}
%$^c$ Since these measurements are in conflict with one another we inflated the uncertainty in the mean by a scale factor $\sqrt{\chi^2/2} = 3.22$.
 \end{flushleft}
 \end{table}

We list some recent evaluations of $M^{0\nu}$ for $^{136}$Xe $\beta\beta$ decay in Table~\ref{tab:NMEs}. While some of these results are in reasonable agreement with each other, there still exist large discrepancies in the calculated values, depending on the method used. Needless to say, this is a pressing issue as future $^{136}{\rm Xe}$ $0\nu\beta\beta$ decay experiments aim to improve their sensitivity by at least one order of magnitude~\cite{nexo:2018}. Additionally, next generation experiments also intend to use the method of barium ion tagging~\cite{McDonald} in xenon time projection chambers (TPCs). This technique has the potential to reduce room background contributions to insignificant levels, which undoubtedly would further enhance the sensitivity of $^{136}$Xe $0\nu\beta\beta$ decay experiments over other experimental searches. 

Finally, it is expected that due to these advantages, future $^{136}$Xe $\beta\beta$ decay experiments would also place the tightest constraints on possible CP-violating Majorana phases in the PMNS matrix~\cite{Pascoli}. The phases in the neutrino mixing matrix are potential leptogenesis parameters that may explain the observed baryon asymmetry in the universe~\cite{Fukugita,Pascoli2,Moffat2019}. 

Quite naturally, due to the several reasons listed above, the $0\nu\beta\beta$ decay of $^{136}$Xe presents a compelling case to address the accuracy in its calculated NME .
%We briefly discuss some salient features of the calculations listed in Table~\ref{tab:NMEs}. 
Two methods that have been traditionally used to calculate $0\nu\beta\beta$ decay NMEs are the interacting shell model (ISM) and the quasiparticle random phase approximation (QRPA). Unlike the latter, the shell model calculations use a limited configuration space that is comprised of relatively fewer single-particle states in the vicinity of the Fermi surface. Despite this restriction, large-scale ISM calculations allow arbitrarily complex correlations between the valence nucleons. On the other hand, the QRPA calculations make use of a much larger model space with comparatively simpler configurations. In general, the ISM calculations are known to yield smaller values of $M^{0\nu}$, compared to the QRPA~\cite{Avignone_rmp,Engel_2017}. This discrepancy has been attributed to different approaches in treating the pairing (or seniority) structure of the nuclear wavefunctions~\cite{ism:2009,Caurier:08}. For the most part, previous QRPA calculations assumed spherical ground states for the parent and daughter nuclei, wherein the pairing correlations between like nucleons were taken into account using the Bardeen-Cooper-Schrieffer (BCS) approximation~\cite{Avignone_rmp,Engel_2017}. It was only recently that deformed QRPA NME calculations were performed for $0\nu\beta\beta$ decays~\cite{qrpa_CH,qrpa:2018}, whose results for $^{136}$Xe are listed in Table~\ref{tab:NMEs}. Compared to the spherical QRPA~\cite{jyvaskyla:2015}, the deformed calculations yield smaller values for the NME, and are in reasonable agreement with the ISM results. The authors of Refs.~\cite{qrpa_CH,qrpa:2018} point out that the suppression of the NME in their calculations is mainly due to differences in the pairing content of the initial and final mean fields. Unlike the spherical QRPA, the deformed calculations accounted for the sharp neutron Fermi surface in $^{136}$Xe due to the neutron number $N = 82$ shell closure. This curtails the overlap between the BCS wavefunctions and leads to a significant reduction in the calculated NMEs~\cite{qrpa_CH,qrpa:2018}. The calculations also suggest that the NME can be even more suppressed if the parent and daughter nuclei have different deformations. Such a scenario will either further reduce~\cite{qrpa_CH,qrpa:2018} the QRPA overlap factors mentioned above or result in a similar seniority mismatch between the ISM wavefunctions, due to high-seniority~\footnote{The seniority quantum number labels the number of unpaired nucleon states in a nucleus~\cite{ism:2009,Caurier:08}} components introduced by the deformation. In comparison, the NME calculations using other many-body approaches such as the non-relativistic energy density functional (NREDF) theory, covariant density functional theory (CDFT), the interacting boson model (IBM-2) or the generator coordinate method (GCM), predict higher values for the NME (Table~\ref{tab:NMEs}). It has been suggested that these values are most likely overestimated, because of the omission of both collective as well as non-collective correlations, depending on the calculation~\cite{gcm:2018,Engel_2017,Menendez:2014,Menendez:2016}. 

In light of the above, precise experimental information elucidating the properties of $^{136}$Xe and $^{136}$Ba nuclei are crucial to benchmark the NME calculations and further reduce their model dependence. Indeed, differences in the valence nucleon occupancies for these nuclei were recently determined using one nucleon transfer reactions~\cite{Entwisle,Szwec}. Furthermore, the ground state of $^{136}$Ba is not expected to have a nearly spherical structure as $^{136}$Xe or $^{138}$Ba. The even barium isotopes in the $N \le 82$ region are known to be transitional, displaying a structural evolution from spherical to $\gamma$-soft behavior with decreasing neutron number~\cite{rmp_deform,Kusakari,Pascu:2010}.
%Recent investigations have shown the low-lying states in $^{136}$Ba to be vibrational in character~\cite{Mukhopadhyay}. 
\begin{figure}[t]
\includegraphics[scale=0.3]{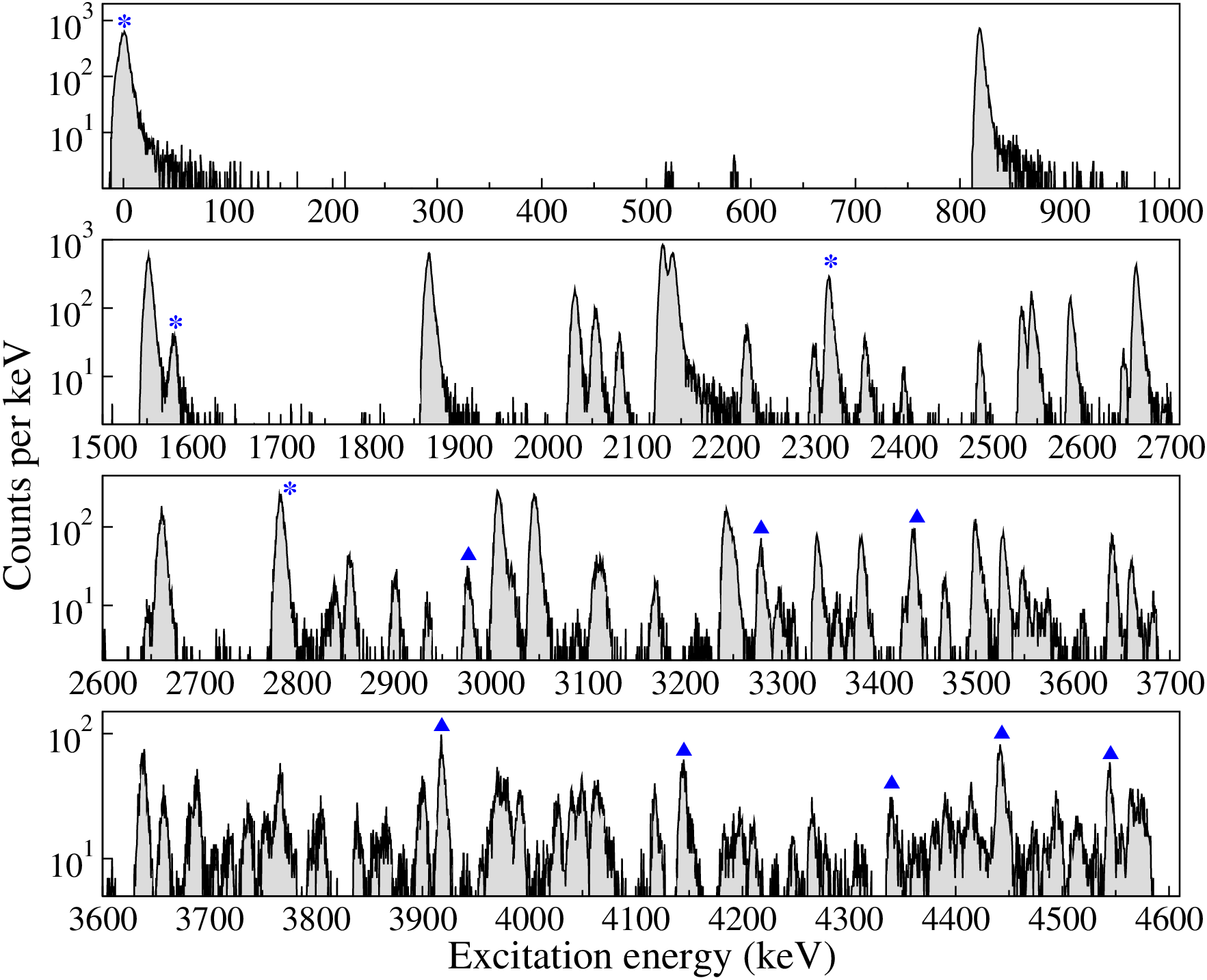}% Here is how to import EPS art
\caption{\label{fig:spectrum}Excitation energy spectrum in $^{136}$Ba obtained at $\theta_{\rm lab} = 25^{\circ}$. Previously known $0^+$ states are marked with asterisks, while the newly identified ones from this work are shown with filled triangles.}
\end{figure}
\begin{figure}[b]
\includegraphics[scale=0.4]{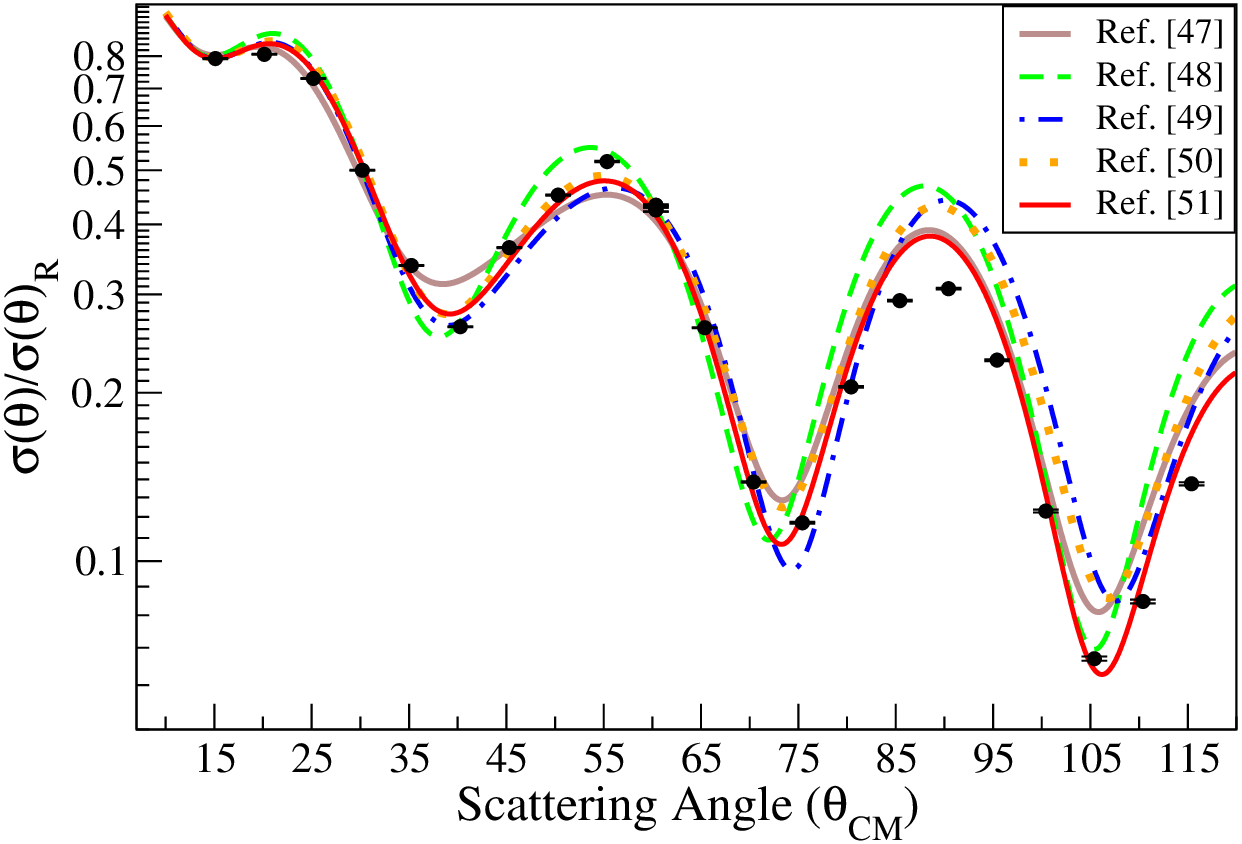}% Here is how to import EPS art
\caption{\label{fig:pp}Measured $^{138}{\rm Ba}(p,p)$ angular distribution from this work (expressed in terms of ratio to the Rutherford cross sections) compared to various DWBA predictions based on different global OMP parameters.}
%~\cite{BG,KD,Menet,WG}.}
%The DWBA analysis assumes a $2n$ pick up configuration of $(0h_{11/2})^2$~\cite{Szwec}.}
\end{figure}
In this Letter we discuss neutron pairing correlations in $^{136}$Ba, studied with the $^{138}{\rm Ba}(p,t)$ reaction. The experiment was performed at the Maier-Leibnitz-Laboratorium (MLL) in Garching, Germany, where a $1.5~\mu$A, 23~MeV proton beam from the MLL tandem accelerator was incident onto a~$40\mu$g/cm$^2$~thick, 99.9\% isotopically enriched $^{138}\rm{BaO}$ target, that was evaporated on a 30~$\mu$g/cm$^2$ carbon backing.
%that had a nominal areal density of $40~\mu$g/cm$^2$ on a 30~$\mu$g/cm$^2$ carbon backing
The light reaction ejectiles were momentum analyzed using the high-resolution Q3D magnetic spectrograph~\cite{Loffler:1973,npn_q3d}, whose solid angle acceptance ranged from $2.3-14.6$~msr during various stages of the experiment. The detector placed at the focal plane of the Q3D spectrograph consisted of two gas proportional counters and one 7-mm-thick plastic scintillator~\cite{npn_q3d}. A cathode strip foil in the second proportional counter provided position information (with a resolution of $\sim 0.1$~mm), while the energy losses in the gas counters and the residual energy deposited in the scintillator allowed for particle identification. The integrated beam intensity was determined using a Faraday cup that was placed at $0^\circ$ to the beam, and connected to a Brookhaven Instruments Corporation (BIC) current integrator.  

\begin{figure}[t]
\includegraphics[scale=0.38]{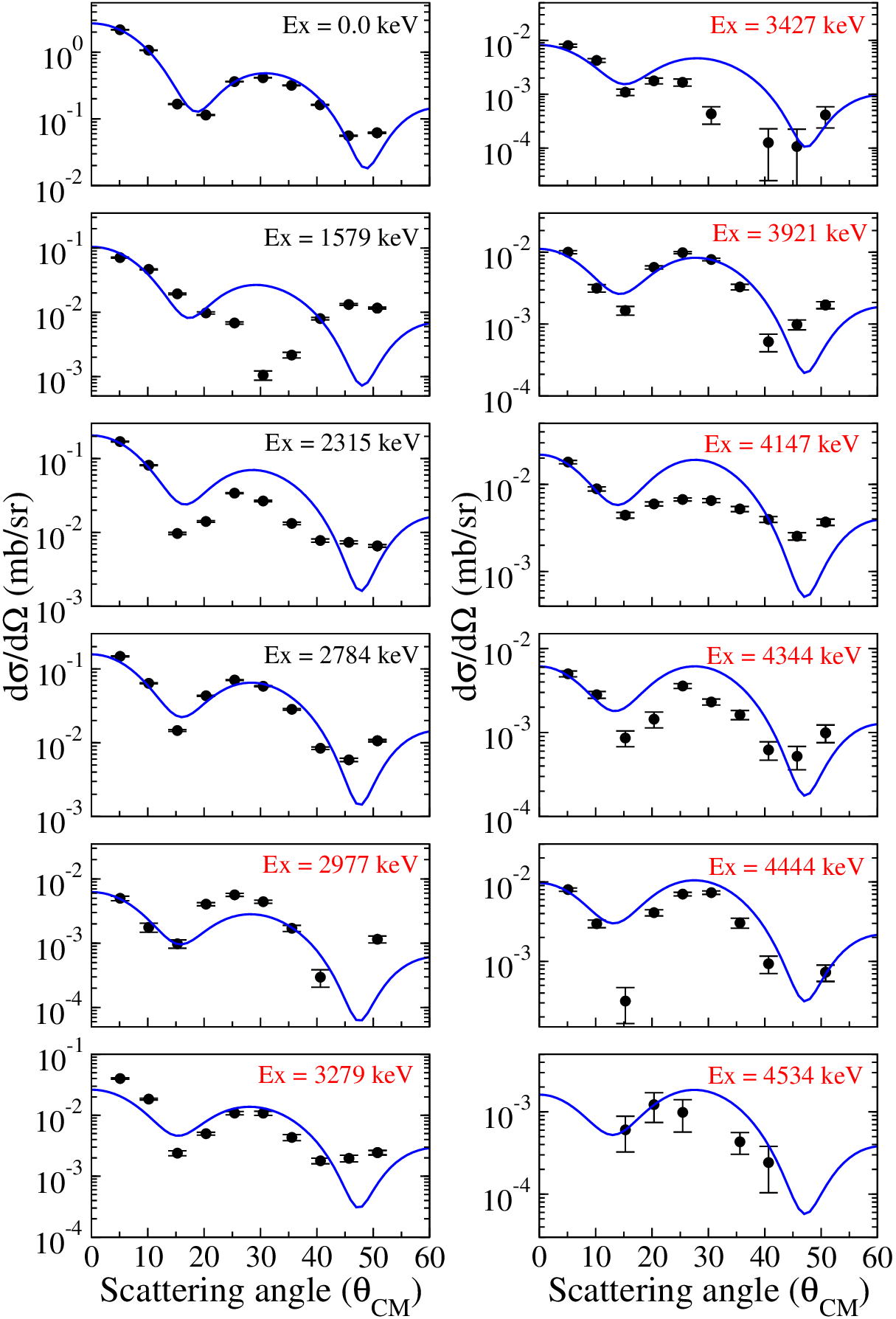}% Here is how to import EPS art
\caption{\label{fig:xsections}Angular distributions of $0^+$ states identified in this work. The solid curves are normalized DWUCK4 DWBA predictions for $L = 0$ transfer, assuming a $(0h_{11/2})^2$ configuration~\cite{Szwec} for the form factor. The newly identified $0^+$ states observed in this work are labeled in red.}
%The DWBA analysis assumes a $2n$ pick up configuration of $(0h_{11/2})^2$~\cite{Szwec}.}
\end{figure}

For our measurements, we obtained triton angular distributions using four magnetic field settings and at ten spectrograph angle settings, ranging from $\theta_{\rm lab} = 5^\circ$ to $50^\circ$. Fig.~\ref{fig:spectrum} shows sample triton spectra obtained from this experiment, where we observe states in $^{136}$Ba up to $\sim 4.6$~MeV in excitation energy. The energy resolution of the triton peaks were found to be $\lesssim$~10~keV. We also took additional $^{138}{\rm Ba}(p,p)$ elastic scattering data over an angular range of $\theta_{\rm lab} = 15^\circ$ to $115^\circ$, in $5^\circ$ steps. These data were used to determine both the effective areal density of the $^{138}$Ba target nuclei, as well as the appropriate global optical model potential (OMP) parameters for the incoming $^{138}{\rm Ba}+p$ reaction channel~\cite{BG,KD,Menet,WG, Varner1991}. The latter were used in a zero-range distorted wave Born approximation (DWBA) analysis of our data, for which we used the DWUCK4 code~\cite{DWUCK4} with Woods-Saxon potentials. As shown in Fig.~\ref{fig:pp}, based on a comparison of various DWBA calculations with our elastic scattering measurements, we chose the global proton OMP parameters recommended by Varner~\textit{et al.}~\cite{Varner1991} for the incoming (proton) channel. For the outgoing $^{136}{\rm Ba}+t$ channel we used the OMP parameters provided by Li~\textit{et al.}~\cite{Li:2007}, as they gave the best agreement with our measured triton angular distribution for the ground state in $^{136}$Ba (c.f. Fig.~\ref{fig:xsections}).
%due to the unavailability of relevant triton elastic scattering data)
 The transfer form factor was calculated assuming a single-step pick up of a di-neutron in a singlet relative $s$-state. For the core-$2n$ coupling we used the global OMP from Ref.~\cite{B_G:1969}, whose well depth was adjusted to reproduce the binding energy of each neutron~\cite{Deason:1979}.
 %$(S_{2n}+E_x)/2$
 %~\cite{Deason:1979}. 
 %
%
 %from the $(0h_{11/2})^2$ configuration in $^{136}$Ba~\cite{Szwec}.
 
 The above approach was used to perform a comprehensive analysis of the angular distributions for all the triton peaks shown in Fig.~\ref{fig:spectrum}. We defer a detailed discussion on the analysis to a forthcoming article~\cite{rebeiro_prc}. In this Letter we only focus on the $L = 0$ angular distributions, which are critical for studying pair-correlated states in even-even nuclei. Our measured angular distributions identify eight new $0^+$ states in $^{136}$Ba~\cite{nds:2018,notes}. These results are shown in Fig.~\ref{fig:xsections}, together with normalized DWBA cross sections. The data show reasonable agreement with DWUCK4 predictions, except for the well-established $0_2^+$ state at 1579~keV~\cite{nds:2018}, where the first minimum occurs at approximately twice the predicted value. This is due to an inherent shortcoming of the DWUCK4 calculations. For example, the calculations ignore multi-step processes and interference from different configurations to the pair transfer amplitude, which can alter the shape of the angular distribution.      
\begin{table}[t]
\begin{flushleft}
\caption{Measured cross sections at $\theta_{\rm lab} = 5^\circ$ and %fragmentation of the $L = 0$ transfer strength relative to the ground state.
relative $(p,t)$ strengths obtained from the data shown in Fig.~\ref{fig:xsections}. 
}
\label{tab:pt_strengths}
\begin{ruledtabular}
\begin{tabular}{l..}
%\multicolumn{1}{c}{$E_x~{\rm (keV)}$}&\multicolumn{1}{c}{Relative} &\multicolumn{1}{c}{$\epsilon$}\\
\multicolumn{1}{l}{$E_x$}&\multicolumn{1}{r}{$(d\sigma/d\Omega)_{5^{\circ}}$}&\multicolumn{1}{c}{$\epsilon_i$}\\
\multicolumn{1}{l}{\rm [keV]}&\multicolumn{1}{r}{[mb/sr]}&\multicolumn{1}{c}{[\%]}\\
 \colrule
    0       &2.17(12)   & 100.0  \\
    1579    &0.071(4)   & 5.1(7) \\ 
    2315    &0.17(1)  & 15.2(19) \\ 
    2784    &0.148(8)  & 14.6(17) \\ 
    2977    &0.0046(6) & 0.65(9) \\ 
    3279    &0.041(2)  & 3.3(3) \\ 
    3427    &0.0082(8) & 1.1(1) \\
    3921    &0.0096(8) & 2.2(3) \\
    4147    &0.018(1) & 5.4(7) \\ 
    4344    &0.0055(6) & 1.8(3) \\
    4444    &0.0075(7) & 3.2(4) \\ 
    4534$^a$    & ...       & 0.6(3)\\ 
\colrule
 \multicolumn{2}{l}{Integrated $L = 0$ strength}& \\
 \multicolumn{2}{l}{relative to the ground state }& \multicolumn{1}{r}{$\sum\epsilon_i$ = 53(3)\%}
\end{tabular}
$^a$ We could not determine the cross section for the $4534$~keV state at low angles due to the presence of a kinematically broadened light-ion contaminant peak in the region.\\ 
%$^b$ In comparison, if we only use the $5^\circ$ data point for the normalizations, the relative $(p,t)$ strength to the excited $0^+$ states is found to be 60(4)\%.    
\end{ruledtabular}
 \end{flushleft}
 \end{table}
In Table~\ref{tab:pt_strengths} we list the measured absolute cross sections for these states at $\theta_{\rm lab} = 5^\circ$, in addition to the $L = 0$ transfer strengths to the excited $0^+$ states relative to the ground state, denoted by $\epsilon_i$. The latter were determined for each excited state by the product
\begin{equation}
 \epsilon_i = \left[
 \frac{ \left(\frac{d\sigma}{d\Omega}\right)^{\rm data}_{0^+{\rm ex}}}{\left(\frac{d\sigma}{d\Omega}\right)^{\rm DWBA}_{0^+{\rm ex}}}
 \right]_i 
 \left[
 \frac{ \left(\frac{d\sigma}{d\Omega}\right)^{\rm DWBA}_{{\rm G.S.}}}{\left(\frac{d\sigma}{d\Omega}\right)^{\rm data}_{{\rm G.S.}}}
 \right]~,
\end{equation}
which was obtained by normalizing the DWBA calculations to the data at forward angles (where the DWBA is best satisfied) for each plot in Fig.~\ref{fig:xsections} and then taking the ratio of the normalization factor for the $0^+_i$ state to the ground state normalization factor. This effectively removes the $Q$-value dependence and other kinematic effects in determining the relative $(p,t)$ strengths. 

Our results in Fig.~\ref{fig:xsections} and Table~\ref{tab:pt_strengths} show that in addition to a number of hitherto unknown $0^+$ states in $^{136}$Ba, we observe a large fragmentation of the $L = 0,~(p,t)$ strength to these states, with $\sim$~30\% of the ground state strength concentrated at 2.3 and 2.8~MeV. This manifestly indicates a breakdown of the BCS approximation for neutrons in $^{136}$Ba~\cite{Freeman:2012}. A similarly large breakdown was not observed in the $^{128-134}$Ba isotopes~\cite{Pascu:2010,Pascu:2009,Suliman:2008}. Nevertheless, such a departure from superfluid behavior is not unexpected in a shape transitional region, particularly around closed shell nuclei~\cite{Freeman:2012,broglia:73}. Therefore, our results indicate that the ground state wavefunctions of $^{138}$$^,$$^{136}$Ba could be largely dissimilar due to the `non-spherical' nature of the latter. Additionally, the data also present evidence of a small pairing gap in $^{136}$Ba, that could occur due to vibrational modes in the pairing field~\cite{broglia:73,brink_book}. While both possibilities cannot be ruled out, the former will have  important ramifications for $^{136}$Xe $0\nu\beta\beta$ decay NME calculations. Previous work showed that the static quadrupole moment of the first $2^+$ state in $^{136}$Ba could be as large as $-0.19$ or $+0.25$~$e$b~\cite{kerns,bechara,rothschild}, which does not rule out a significantly deformed $^{136}$Ba ground state.  
 
The results from this experiment also allow tests of the nuclear structure models that are used to calculate the NME for $^{136}$Xe $0\nu\beta\beta$ decay. We performed one such test using configuration interaction shell model calculations with the NuShellX code~\cite{nushellx}. Unlike the DWUCK4 calculations which merely served to identify the observed $0^+$ states and determine relative strengths, the shell model calculations were used to assess the absolute $(p,t)$ cross sections, distributed over $0^+$ states in $^{136}$Ba. The calculations used the five-orbital $(0g_{7/2}, 1d_{5/2}, 1d_{3/2}, 2s_{1/2}, 0h_{11/2})$ valence space for protons and neutrons to determine the wavefunctions for the 0$^{ + }$ ground state of $^{138}$Ba and the lowest fifty 0$^{ + }$ states in $^{136}$Ba. NuShellX was then used to calculate two-neutron transfer amplitudes between these states, that served as input for the Fresco~\cite{fresco} coupled-reaction channels code to generate $^{138}{\rm Ba}(p,t)$ angular distribution predictions. In the Fresco calculations we used the same OMP parameters as the DWUCK4 calculations (for the proton and triton channels) and took into account the coherent sum of both direct and sequential two-step transfer. The single nucleon transfer amplitudes for the two-step part were obtained assuming one intermediate state in $^{137}$Ba for each of the transferred $(n,\ell,j)$ values. For the $^{137}{\rm Ba}$-deuteron coupling, we used the global OMP parameters recommended by An and Cai~\cite{An}, based on a comparison with $^{138}{\rm Ba}(d,d)$ angular distribution data (similar to Fig.~\ref{fig:pp}) that we obtained independently. 
\begin{figure}[t]
\includegraphics[scale=0.33]{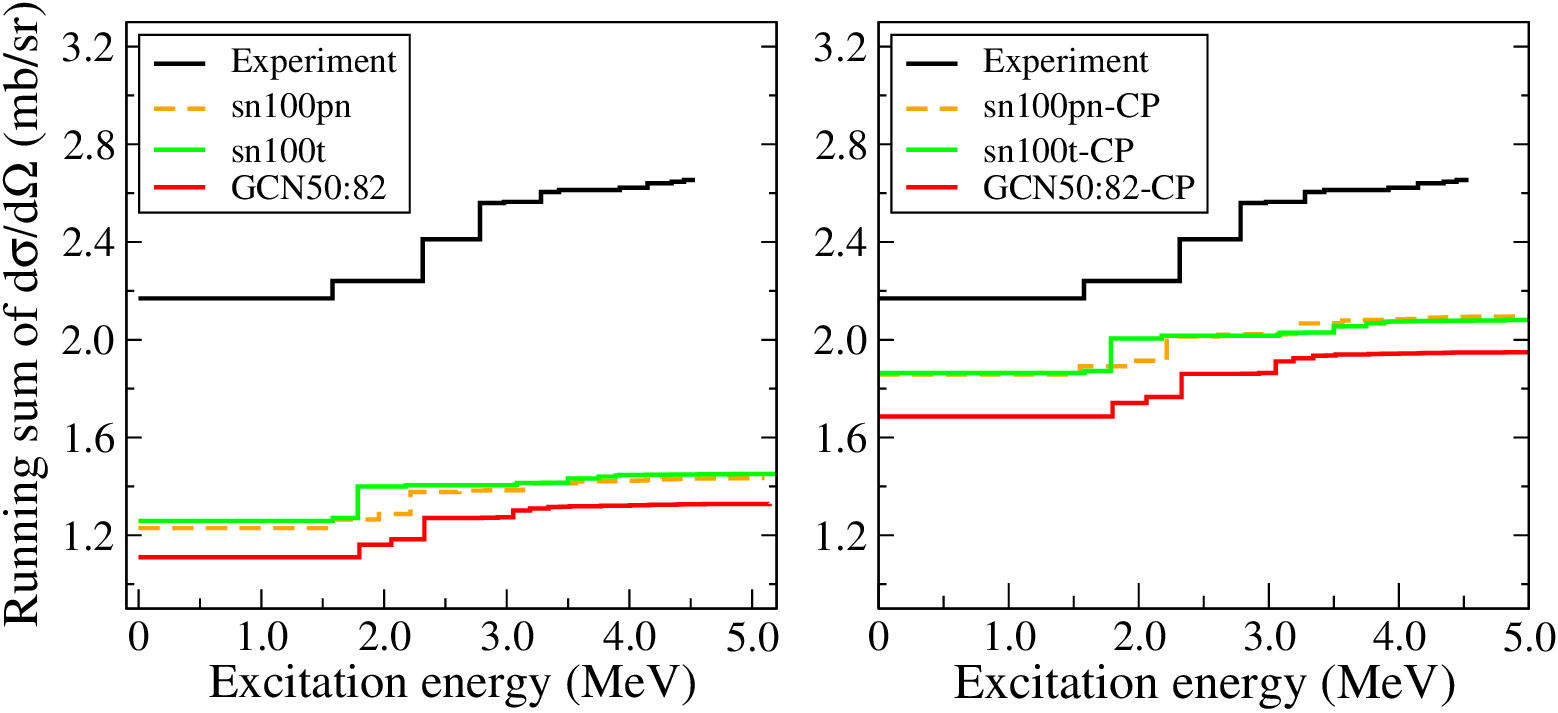}% Here is how to import EPS art
\caption{\label{fig:sum}A comparison of the running sum of experimental $(p,t)$ cross sections at $\theta_{\rm lab} = 5^\circ$ with the calculations described in the text. Left panel: The calculated values without core-polarization corrections to the two-neutron transfer amplitude. Right panel: Results after core-polarization effects are taken into account.}
%The DWBA analysis assumes a $2n$ pick up configuration of $(0h_{11/2})^2$~\cite{Szwec}.}
\end{figure}
We used three different Hamiltonians for the calculations, which were corrected for core-polarization due to configuration mixing with orbitals outside the model space~\cite{gmatrix}.  
% whose single-particle energies were adjusted to reproduce single proton states in $^{133}$Sb and single neutron (hole) states in $^{131}$Sn.  
% 
% 
% The single-particle energy for all of these were adjusted to give experimental spectra relative
% to the closed shell of $^{132}$Sn. 
The first Hamiltonian is from Ref.~\cite{Brown:05} and is called sn100pn in the NuShellX interaction library \cite{nushellx}. 
% This Hamiltonian contains different  
% neutron-neutron $(nn)$, proton-neutron $(pn)$ and proton-proton $(pp)$ two-body matrix elements (TBME) for isospin $T=1$, with the Coulomb interaction included for the $pp$ component. The TBME were obtained from the Brueckner $G$~matrix~\cite{gmatrix} elements of the Paris potential~\cite{paris1,paris2}, and corrected for core-polarization due to configuration mixing with orbitals outside the model space~\cite{gmatrix}. 
The second Hamiltonian (that we call sn100t) is very similar to sn100pn, except with minor modifications and
%to conserve isospin and agree with experimental spectra
was used in Ref.~\cite{Horoi:2013} to calculate $M^{0\nu}$ for $^{136}$Xe $\beta\beta$ decay, while the third GCN50:82~\cite{gcn} Hamiltonian was used by the authors of Ref.~\cite{Caurier:08} to calculate the NME for the decay.
% It is essentially the same as sn100pn, except that it was modified to conserve isospin 
% %by replacing the $pp$ and $pn$ TBME with the $nn$ TBME, 
% and required minor adjustments 
% in the monopole part of the interaction to better agree with experimental spectra for $^{133}$Sb and $^{131}$Sn. 
% For the third calculation we used the GCN50:82~\cite{gcn} isospin conserving Hamiltonian 
% that was also obtained from the $G$-matrix and based upon a realistic CD-Bonn potential~\cite{cdbonn}. 

In the left panel of Fig.~\ref{fig:sum} we compare the running sum of our measured $(p,t)$ cross sections at the most forward angle (where the $L = 0$ strength is concentrated) to the shell model/Fresco calculations. Similar to our experimental results, the calculations predict the cross section to be dominated by the transition to the ground state in $^{136}$Ba, with smaller contributions from excited $0^+$ states. However, the theory predictions are found to be about a factor of two smaller than the experimental values. Such an underestimation should not be surprising, given that the model space for neutrons is limited to only five orbitals near the Fermi surface. Coherent contributions from all orbitals outside the valence space are known to enhance the calculated $L = 0$ two-neutron transfer cross section~\cite{decowski}. We next considered the effects of such core-polarization by calculating ladder-diagram corrections to the two-nucleon transfer amplitudes (TNA), as described in Ref.~\cite{decowski}, 
%. These calculations 
assuming the scattering of pairs of neutrons to twenty three orbitals beyond the model space (up to $i_{11/2}$). 
%and used the single-particle energies for $^{132}$Sn obtained with the Skx Skyrme functional~\cite{skx}. The M3Y potential~\cite{Bertsch,love} was used for the residual interaction. 
As shown in the right panel of Fig.~\ref{fig:sum}, the revised calculations that incorporated the core-polarization effects show enhancements in the predicted cross sections by about a factor of $1.5$, and agree with experiment to $\sim 22$\% for the GCN50:82 Hamiltonian and $\sim 14\%$ for the others. It is worthy of note that the relative distribution of the predicted cross sections over $0^+$ states also agree reasonably well with experiment, particularly for the GCN50:82
 Hamiltonian. 
The agreement did not significantly improve on making small adjustments of the single-particle energies and pairing strengths of the Hamiltonians.
%On scaling the calculations to match the measured ground state cross section, we observe that the result obtained using the GCN50:82 Hamiltonian best agrees with our experiment (c.f. Fig.~\ref{fig:sum}). 

How does the above relate to NME calculations for $^{136}$Xe $0\nu\beta\beta$ decay? 
The connection is discussed in Ref.~\cite{br14}, where it was shown that the $0\nu\beta\beta$ decay NME for a parent nucleus with mass number $A$ can be expanded as a sum over states in an intermediate nucleus with mass number $(A - 2)$. 
%in the case of $^{76}$Ge $\overset{\beta\beta}{\to}$ $^{76}$Se, whose NME was expanded as a sum over intermediate states in the $(A - 2)$ nucleus $^{74}$Ge~\cite{br14}.
For the case of $^{136}$Xe, we can similarly evaluate the NME by summing over the products of the TNA for two-neutron removal to $^{134}$Xe, the TNA for two-proton addition to $^{136}$Ba, and the two-body matrix element for the double-beta decay operators (c.f. Eq.~9 in Ref.~\cite{br14}). The most significant contribution to the NME is through the $0^+$ ground state in the $^{134}$Xe, while $J > 0$ intermediate states mainly cancel the $\Delta J = 0$ term~\cite{br14}. This is similar to other calculations~\cite{Caurier:08,Menendez:18} that separate the NME in terms of nucleon pairs coupled to angular momentum and parity $J^\pi = 0^+$ and $J^\pi \ne 0^+$, where the $J > 0$ contributions predominantly cancel the leading $J^\pi = 0^+$ term (see Fig.~1 in Ref.~\cite{Caurier:08}).

Since the \mbox{$^{136}{\rm Xe}$ $\to$ $^{134}{\rm Xe}$} transition is similar to \mbox{$^{138}{\rm Ba}$ $\to$ $^{136}{\rm Ba}$} (they both correspond to a  transformation from $N = 82$ to $N =80$), we can benchmark the dominant $J^\pi = 0^+$ Gamow-Teller (GT) component of the NME, by calculating it using the expanded set of TNA that better reproduces our measured $^{138}{\rm Ba}(p,t)$ cross section. To do so, we first performed a five-orbital valence space ISM calculation of this part of the NME (for light neutrino exchange) with the sn100t Hamiltonian. On using the CD-Bonn potential~\cite{cdbonn} for two-nucleon short range correlations (SRC) and further including higher-order contributions (HOC) due to induced nucleon currents~\cite{hoc_paper}, we determine the matrix element to be $M_{GT}^{0\nu} (J^\pi = 0^+) = 5.67$.
%~\footnote{This calculation assumes light neutrino exchange, and neglects short range correlations (SRC) and higher order contributions (HOC) to the nuclear current}.
Next we evaluated the NME through the $J^\pi = 0^+$ ground state in $^{134}$Xe, both with and without the core-polarization corrections to the TNA described above. These calculations showed that the five-orbital valence space ISM result for the NME required an enhancement factor of $\sim 1.58$, due to the required core-polarization corrections. Such an enhancement significantly increases the magnitude of the NME, so that its revised value is $M_{GT}^{0\nu} (J^\pi = 0^+) = 8.96$.    
%In comparison, evaluations of the NME, performed both with and without core-polarization corrections to the TNA, show that $M_{GT}^{0\nu} (J^\pi = 0^+)$ requires an enhancement factor of $\sim 1.58$, to reconcile with our experimental result.    
%On using the GCN50:82 Hamiltonian this yields $M_{GT}^{0\nu} (J^\pi = 0^+) = xxx$, for light neutrino exchange, which is a factor of $\sim 1.54$ larger than the NME evaluated with no core-polarization corrections to the TNA or a previous five-orbital valence space ISM calculation $M_{GT} (J^\pi = 0^+) = 5.72$~\cite{Caurier:08}. 
To make further comparisons, we also performed a large-scale spherical QRPA calculation of the NME (with HOC and the CD-Bonn SRC), using the model parameters of Ref.~\cite{Simkovic} and 28 orbitals for major oscillator shells with $N \le 6$. This resulted in a value $M_{GT}^{0\nu}(J^\pi = 0^+) = 9.63$. It is indeed gratifying that the QRPA result (from using a significantly larger number of orbitals) is closer in value to our revised calculation of the NME, compared to the five-orbital model space result. Clearly, previous configuration interaction (ISM) calculations~\cite{Caurier:08} had underestimated the $J = 0$ component of the Gamow-Teller NME. We recommend improved calculations of this part of NME as well as the canceling $M_{GT}^{0\nu} (J > 0)$ term, perhaps along the lines of a many-body perturbation theory treatment~\cite{Holt:13}, that takes into account physics contributions from beyond the model space. Since the details of the cancellation between the $J = 0$ and $J > 0$ contributions to the NME are model dependent, our result also serves to benchmark future calculations of $M_{GT}^{0\nu}(J^\pi = 0^+)$ along these lines. We note that in order to make the connection with two-nucleon transfer reaction data, it is important that the results of these calculations be expressed in terms of their $  J^\pi  $ decomposition.

In summary, this work demonstrates for the first time a direct evaluation of part of a $0\nu\beta\beta$ decay NME using experimental data. In addition to providing a benchmark for future calculations, it also presents a new avenue of approach for evaluating $0\nu\beta\beta$ decay NMEs more accurately, motivating similar investigations in other candidates.  We also report for the first time a large breakdown of the neutron BCS approximation in an even barium nucleus with $N \le 82$. Our observations motivate a reassessment of the neutron pairing approximation in $^{136}$Ba, used in some NME calculations for $^{136}$Xe $0\nu\beta\beta$ decay and invite further investigations of the shape of $^{136}$Ba.

We are grateful to Ian Thompson for his assistance with the Fresco calculations. This work was partially supported by the National Research Foundation (NRF) of South Africa, the Natural Sciences and Engineering Research Council of Canada (NSERC), the U.S. National Science Foundation under Grant No.~PHY-1811855 and the U.S. Department of Energy, Office of Science under Grant No.~DE-SC0017649. P.A. acknowledges funding from the Claude Leon Foundation in the form of a postdoctoral fellowship.

\bibliography{138ba_pt}% Produces the bibliography via BibTeX.

\end{document}